\documentclass[aip,jcp,groupedaddress,a4paper,12pt,reprint]{revtex4-1} 
		\usepackage[utf8]{inputenc} 
		\usepackage[T2A, T1]{fontenc} 
		\usepackage{lmodern} 
		  \usepackage[english]{babel}
   \usepackage{amsmath}   %
   \usepackage{amsfonts}  
   \usepackage{amssymb}   %
   \usepackage{graphicx}  
	\usepackage{array} 
	\usepackage{multirow} 
	\usepackage{rotating} 
   \usepackage{multirow} 

\begin{document} 

\title{DNA-stabilized Ag--Au bimetallic clusters: The effects of alloying and embedding on optical properties}


\author{Dennis Palagin}
\email{dennis.palagin@chem.ox.ac.uk}
\affiliation{Physical and Theoretical Chemistry Laboratory, Department of Chemistry, University of Oxford, South Parks Road, Oxford, OX1 3QZ, United Kingdom}

\author{Jonathan P. K. Doye}
\affiliation{Physical and Theoretical Chemistry Laboratory, Department of Chemistry, University of Oxford, South Parks Road, Oxford, OX1 3QZ, United Kingdom}

\begin{abstract}
Global geometry optimization and time-dependent density functional theory calculations have been used to study the structural evolution and optical properties of Ag$_{n}$Au$_{n}$ ($n =$ 2--6) nanoalloys both as individual clusters and as clusters stabilized with the fragments of DNA of different size. We show that alloying can be used to control and tune the level of interaction between the metal atoms of the cluster and the organic fragments of the DNA ligands. For instance, gold and silver atoms are shown to exhibit synergistic effects in the process of charge transfer from the nucleobase to the cluster, with the silver atoms directly connected to the nitrogen atoms of cytosine increasing their positive partial charge, while their more electronegative neighbouring gold atoms host the excess negative charge. This allows the geometrical structures and optical absorption spectra of small bimetallic clusters to retain many of their main features upon aggregation with relatively large DNA fragments, such as a cytosine-based 9-nucleotide hairpin loop, which suggests a potential synthetic route to such hybrid metal-organic compounds, and opens up the possibility of bringing the unique tunable properties of bimetallic nanoalloys to biological applications.
\end{abstract}

\maketitle

\section{Introduction}

\begin{figure*}[htp!]
\centering
\includegraphics[width=0.67\linewidth]{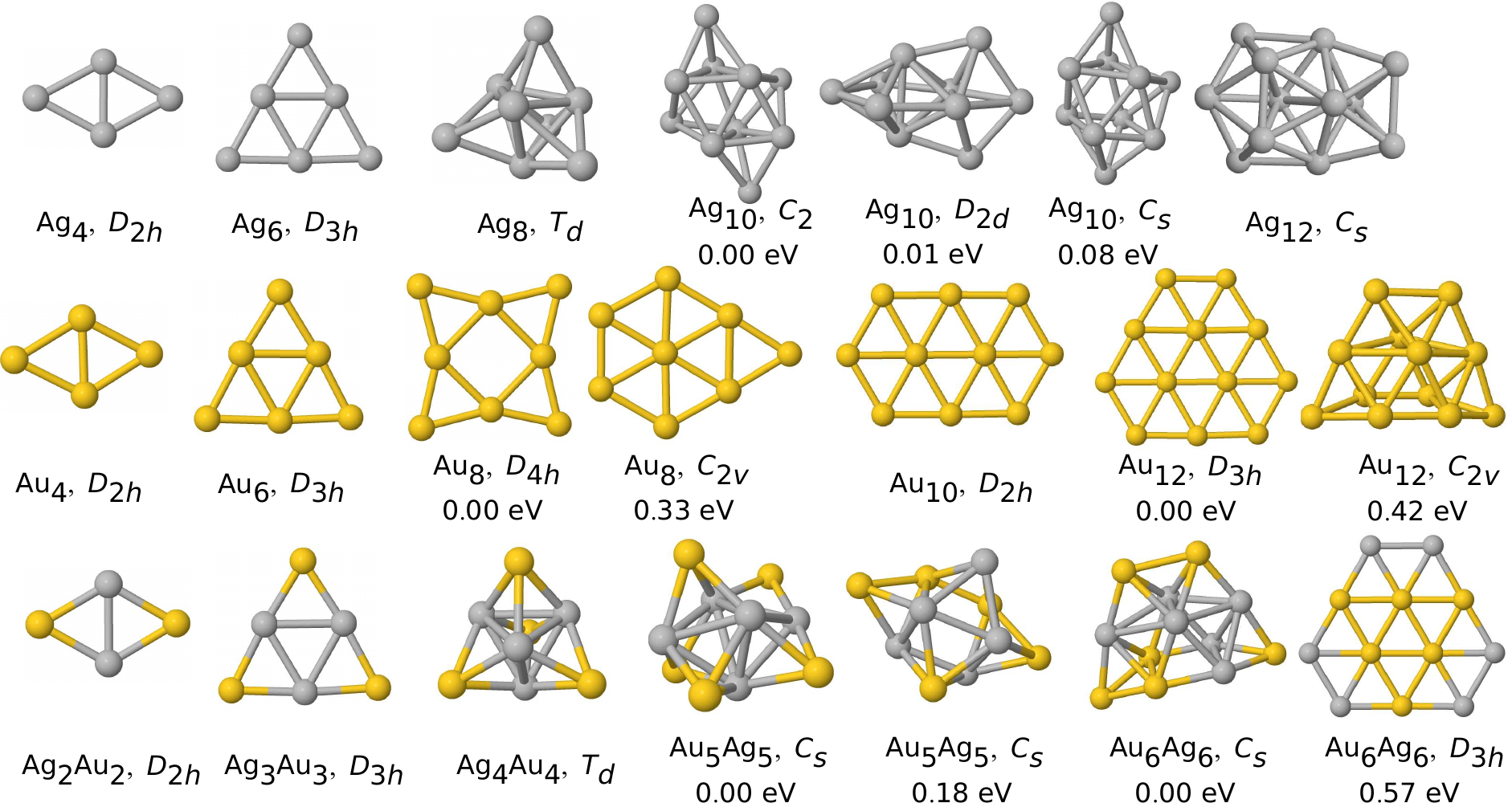}
\caption{Ground-state and selected low-lying isomers of Ag$_{n}$, Au$_{n}$, and Ag$_{n}$Au$_{n}$.}
\label{fig01}
\end{figure*}

Metal clusters are known to exhibit unique optical properties.~\cite{Kreibig_Vollmer_book_1995} Moreover, the properties of the clusters in the small size regime depend not only on composition, but also on their size and shape, which opens the possibility of fine-tuning the optical range of absorption and emission.~\cite{Kelly_Coronado_Zhao_Schatz_2003, Cottancin_Celep_Broyer_2006} Such tunability suggests potential applicability in a wide variety of fields, including imaging, sensing, biology, and medicine.~\cite{Jain_Huang_Sayed_2008, McNamara_perspective} Gold nanoclusters, especially, have attracted much attention due to the rich diversity of their structures and the flexibility of their optical properties.~\cite{Lerme_Broyer_1998, Link_Sayed_2000, Aizpurua_Abajo_2003, Palagin_Matulis_Nozaki_2010, Philip_Thomas_2012, Yu_Xi_2013, Stamplecoskie_2014, Burgess_2014} Optical absorption spectra of small gold clusters have been studied both experimentally and theoretically.~\cite{Omary_2001, Idrobo_2007, Lecoultre_2011, Koppen_2012, Alvino_2013} Silver clusters have also been studied extensively.~\cite{Idrobo_Jellinek_2008, Harb_Felix_2008} As the field of bimetallic nanoalloys has grown,~\cite{Ferrando_ChemRev_2008} it was noted that alloying allows finer control of the optical properties, providing an additional variable to tune the properties of such clusters.~\cite{Gaudry_Melinon_2003, Cortie_McDonagh_2011, Barcaro_Stener_2015} For instance, Ag--Au bimetallic clusters seem promising in this context.~\cite{Zhang_Teo_2001, Li_Wilcoxon_Johnston_2008, Barcaro_Fortunelli_Stener_2011, Longsdail_Johnston_2012, Lopez_Weissker_2013, Shayeghi_Johnston_2014, Ag_Au_emission, ANIE:ANIE201107696, ANIE:ANIE201307480}

However, the ultimate goal is to bring these novel properties to real materials and applications. For this task, such clusters should be stable and able to function in the medium of interest, for example the living cell, be it for bio-labelling,~\cite{Lin_biolabelling} imaging,~\cite{Palmal_2014} medical purposes,~\cite{Shankar_medical, Conde_medical} or as analytical sensors.~\cite{Yang_Vosch_2011, Bagga_perspective} For example, quite an extensive body of recent literature is devoted to the nanostructures of silver clusters adsorbed, onto DNA strands.~\cite{Driehorst_Fygenson_2011, Han_Wang_2012, Oneill_Fygenson_2012, Gwinn_Oneill_2008, Oneill_Velazquez_2009, Schutz_Gwinn_2013, Ramazanov_Kononov_2013, Markesevic_Bouwmeester_2014, Copp_Gwinn_2014, Swasey_Gwinn_2014, Copp_Gwinn_2015, Fortino_Russo_2015, Buceta_2015, doi:10.1021/jp206251v, doi:10.1021/acs.analchem.5b01265, C5RA11271K, doi:10.1021/acs.analchem.5b03166, doi:10.1021/acs.jpcc.5b08834, nano5010180} One of the central questions in an investigation of such hybrid materials is the reliable identification of the possible geometrical configurations of synthetically feasible aggregates; however, on this question no consensus has been reached. On the one hand, DNA oligomers were suggested to be able to stabilize silver clusters of certain shapes, such as ``nanorods'', that are not typical for individual nanoclusters.~\cite{Schutz_Gwinn_2013, Copp_Gwinn_2014, Swasey_Gwinn_2014} Such elongated clusters have been suggested to bear positive charge on the atoms in contact with nucleobases,~\cite{Schutz_Gwinn_2013, Copp_Gwinn_2014} with the possibility of using DNA as a template for the directed synthesis of chains of silver nanoclusters from silver ions to produce a conducting nanowire.~\cite{silver_wires} Theoretical studies have also considered the energetic stability of silver-ion-mediated mismatch base pairs.~\cite{Fortino_Russo_2015} On the other hand, experimental evidence also suggests that compact small clusters, such as Ag$_{3}$, might be intercalated into short duplexes, albeit with significant structural distortion of the double helix.\cite{Buceta_2015} Furthermore, the experimental feasibility of the selective assembly of larger silver clusters (up to at least $\sim10$ atoms) attached to certain specific sites of the DNA molecule, such as hairpins, have been proposed.~\cite{Driehorst_Fygenson_2011, Oneill_Fygenson_2012, doi:10.1021/acs.analchem.5b01265, doi:10.1021/acs.analchem.5b03166, doi:10.1021/acs.jpcc.5b08834} Such selectivity could provide potential routes towards nanotechnological applications, for instance in the field of nano-optics.~\cite{Copp_Gwinn_2015} However, the favourable configurations of the DNA-bound metal clusters, as well as their preferred binding sites, remain an open question.


\vspace{1cm}

These challenges motivate the present theoretical investigation of the influence of DNA bases on the geometry and absorption spectra of pristine Au, Ag, and bimetallic Ag--Au clusters, focusing on clusters with 4, 6, 8, 10, and 12 atoms. Small clusters up to 12 atoms have been chosen for the following reasons. Firstly, the structures of gold and silver clusters in this size range are known, while the data on nanoalloys is limited. Secondly, for these sizes gold clusters prefer planar configurations that are very different from their silver counterparts (for $n > 6$), making the interplay between the two metals of particular interest. Finally, this is the size range relevant to the experiments on gold and silver clusters complexed with DNA.~\cite{Oneill_Fygenson_2012, Buceta_2015, doi:10.1021/acs.analchem.5b01265} 


We therefore systematically investigate the aggregation of these metal clusters with increasingly complex DNA fragments in order to study the possibility of the formation of stable aggregates through, for example, the wrapping of single-stranded sections of DNA, such as hairpins, around a cluster. Besides studying the geometrical configurations of metal clusters, we identify the most suitable nucleobases for cluster stabilization, investigate the chemical nature of the interaction between a cluster and a DNA fragment, and explore how the optical properties of the cluster change with alloying and embedding. Understanding and having the ability to control these spectral changes is crucial for  biotechnological applications such as labelling~\cite{Lin_biolabelling} or sensing.~\cite{Yang_Vosch_2011, Driehorst_Fygenson_2011}

It should be emphasized that the expense of our computational setup only allows us to look at individual aggregates without solvent molecules present in the system, thus setting certain limits for the interpretation of our results for DNA-based nanomaterials. The choice of the solvent might change the optical properties of the stabilized metal nanoparticles by affecting the interaction between a ligand and a cluster. For instance, the pH of the solvent might influence the basicity/acidity of the coordination centers.~\cite{Ag20_solvent} However, as the main optical transitions in hybrid systems are typically due to metal cluster orbitals, theory predicts solvent effects to be minor,~\cite{doi:10.1021/jp9051853, C4CP06103A} as confirmed in Supplementary Information, section SII. We therefore expect to be able to identify the general trends in how the geometrical configurations and optical properties of the metal clusters change upon aggregation with increasingly large DNA fragments, and to describe the nature of the chemical bonding in such structures. These qualitative results are expected to shed light on the possibility of creating stable hybrid metal-organic materials and the tuning of their properties by alloying, with the potential for a wide range of applications.


It should be noted that in our discussion of the metal clusters aggregated with DNA fragments we focus on the interaction of the pre-formed neutral clusters with nucleobases, which can be experimentally achieved, for instance, using the recently proposed  embedding of an electrochemically pre-formed clusters into an individually stable DNA fragment.~\cite{Buceta_2015} The self-assembly of the clusters from metal ions in the presence of DNA, which is typically studied in a AgNO$_{3}$ solution,~\cite{doi:10.1021/acs.analchem.5b01265} goes beyond the scope of this investigation.

\section{Geometries and optical spectra of individual clusters}

Firstly, we ran density functional theory (DFT) based global geometry optimization to find the most stable configurations of individual pristine and alloyed clusters. The identified ground-state structures are presented in Fig.~\ref{fig01}. The geometrical configurations of Ag$_{4}$, Ag$_{6}$, Ag$_{8}$, and Ag$_{12}$ clusters fully agree with the well established structures previously reported in the literature.~\cite{Harb_Felix_2008} For Ag$_{10}$ we have identified three low-energy isomers within 0.1~eV: in addition to the commonly identified $D_{2d}$~\cite{Harb_Felix_2008, Fournier_2001} structure, we also found $C_{s}$ and $C_{2}$ configurations, both of which were reported earlier in different sources.~\cite{Harb_Felix_2008, Hong_AgAu} This is consistent with the experimental evidence suggesting the existence of several Ag$_{10}$ isomers in a narrow energy range.~\cite{Harb_Felix_2008, Conus_Felix_2006} 

The planar geometries of the ground-state structures for the Au$_{4}$, Au$_{6}$, and Au$_{8}$ clusters agree with those previously reported.~\cite{BravoPerez_1999, Hakkinen_2000, Remacle_2005, Palagin_Matulis_Nozaki_2010} For neutral gold clusters larger than Au$_{8}$, there has been an active discussion concerning the cluster size at which the transition from planar to 3D structures occurs.~\cite{Fernandez_Balbas_2004, Li_Au10, Hansen_2013} With the transition suggested to take place around $n=12$ for neutral Au$_{n}$ clusters~\cite{Fernandez_Balbas_2004} and Au$_{n}^{-}$ anions,~\cite{Furche_2002, Johansson_2008} recent theoretical investigations have proposed planar $D_{2h}$~\cite{Fernandez_Balbas_2004, Li_Au10} and  $D_{3h}$~\cite{Fernandez_Balbas_2011} structures as the ground states for Au$_{10}$ and Au$_{12}$, respectively. Our global optimization results also suggest that planar structures are indeed the most stable for both Au$_{10}$ and Au$_{12}$, with the Au$_{12}$ $D_{3h}$ structure being 0.42~eV more stable than the previously suggested~\cite{Fernandez_Balbas_2004} three-dimensional $C_{2v}$ structure.

The structural data on small bimetallic Ag--Au clusters is rather limited. Theoretical studies have been carried out to determine the configurations of singly-doped gold clusters,~\cite{Tafoughalt_2014} 4-atom~\cite{Shayeghi_Johnston_2014} and 8-atom~\cite{Heiles_Johnston_2012} nanoalloys, as well as Ag$_{n}$Au$_{m}$ ($2 \leq n+m \leq 8$~\cite{Zhao_Zeng_2006} and $5 \leq n+m \leq 12$~\cite{Hong_AgAu}) clusters of various composition. The ground-state structures of the Ag$_{2}$Au$_{2}$, Ag$_{3}$Au$_{3}$, and Ag$_{4}$Au$_{4}$ nanoalloys identified by our global optimization agree with those suggested in Refs.~\citenum{Shayeghi_Johnston_2014}, \citenum{Zhao_Zeng_2006}, and \citenum{Heiles_Johnston_2012}, respectively. For Ag$_{5}$Au$_{5}$ we identified a new $C_{s}$ global minimum, which is 0.18~eV more stable than the previously suggested $C_{s}$ structure.~\cite{Hong_AgAu} The cluster structure consists of a silver bipyramidal core whose faces or edges are capped by gold atoms and follows a similar pattern to the Ag$_{4}$Au$_{4}$ tetracapped tetrahedron. For the Ag$_{6}$Au$_{6}$ cluster we identified a new ground-state structure, which lies lower in energy than both previously suggested~\cite{Hong_AgAu} ones: a newly found $C_{s}$ cluster, similar to the second-lowest isomer of Au$_{12}$, is 0.57~eV more stable than the planar $D_{3h}$ minimum, and 0.40~eV more stable than the previously suggested compact $C_{s}$ structure.

How does the alloying influence the geometry of metal clusters? All global minima of the nanoalloys have silver atoms located closer to the center of the cluster in sites with higher coordination number, thus forming a ``core'', while gold atoms surround the silver core, forming a ``shell''. Although individual gold clusters have higher dissociation energies than their silver counterparts, such an arrangement allows both the gold and silver atoms to satisfy their relative preferences for lower and higher coordination environments that is exhibited in the global minima of the pure clusters. Ag$_{2}$Au$_{2}$ and Ag$_{3}$Au$_{3}$ retain the $D_{2h}$ and $D_{3h}$ geometries exhibited by their pristine counterparts. For the larger bimetallic clusters, retaining the planar geometry of the correspondingly sized gold clusters is energetically unfavourable. Instead, Ag$_{4}$Au$_{4}$ adopts the $T_{d}$ configuration of the pure silver cluster with a central 4-atom silver tetrahedron capped on each of its faces by a gold atom. The larger Ag$_{n}$Au$_{n}$ clusters continue this trend of three-dimensional core-shell structures, but the global minima no longer match those for the pure silver clusters.

\begin{figure}[!ht]
\centering
\includegraphics[width=\linewidth]{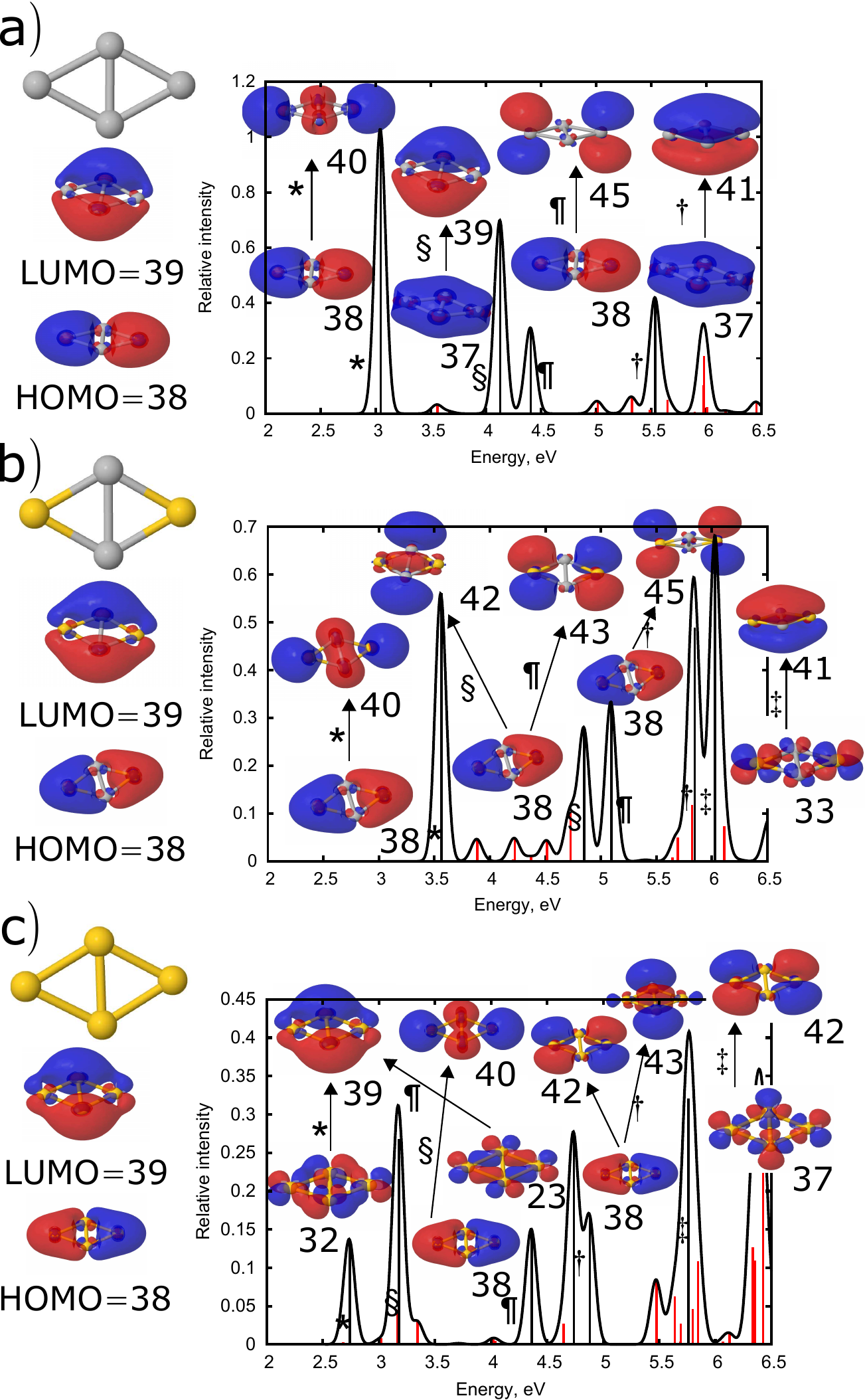}
\caption{Comparison of the excitation transitions for \mbox{(a) Ag$_{4}$}, (b) Ag$_{2}$Au$_{2}$, and (c) Au$_{4}$. Red sticks correspond to the calculated transition energies. Black continuous lines are obtained by Gaussian broadening ($\sigma = 0.05$~eV). Molecular orbitals are depicted as 3D isosurfaces at 0.02~$e/\text{\AA}^3$. The illustrated orbitals correspond to some of the more dominant transitions in the spectrum; the relevant peaks are coloured black and marked with the same symbol. The insets on the left of the spectrum depict the ground-state structure of the cluster, and its highest occupied (HOMO) and lowest unoccupied (LUMO) molecular orbitals. The numbering of orbitals corresponds to the number of explicitly described electron pairs.}
\label{fig02}
\end{figure}

An additional feature of alloying is a partial charge transfer within the cluster from silver atoms to more electronegative gold atoms. While pure gold and silver clusters exhibit only marginal charge redistribution, with the surface atoms carrying a small excessive negative charge (up to about $-0.2$\,$e$), in the case of alloyed clusters such a charge transfer is noticeably intensified, with the partial negative charge on the gold atoms reaching about $-0.5$\,$e$. For instance, in the case of Ag$_{4}$Au$_{4}$ it results in the Au shell having a total negative charge of $-1.72$\,$e$. It has been shown that such charge localization in bimetallic clusters occurs due to the difference in electronegativity of the constituent metals, and can determine relative homotop stabilities and chemical activity of the cluster.~\cite{Johnston_charge_localization, C0CP02845B}

Next, we consider the influence of the alloying on the optical absorption spectra of the clusters. The spectra of the smallest Ag$_{2}$Au$_{2}$ (Fig.~\ref{fig02}) and Ag$_{3}$Au$_{3}$ (see Supplementary Information) clusters, where the nanoalloys and pure clusters have the same geometrical structure, appear to be a mixture of the features seen in the spectra of Ag$_{4}$ and Au$_{4}$, or Ag$_{6}$ and Au$_{6}$, respectively, as they mostly consist of transitions between equivalent orbitals to the pure clusters. The peak positions, however, do not necessarily match exactly due to the effects of alloying on the relative orbital energies. In the case of Ag$_{2}$Au$_{2}$, the lower energy range of the spectrum more resembles that of Ag$_{4}$, while the higher energy range is closer to the Au$_{4}$ spectrum. Consistent with this, the character of the orbitals involved in excitation transitions in the case of the bimetallic cluster are similar to Ag$_{4}$ for low energies, and to Au$_{4}$ for higher energies. A similar trend is also observed for Ag$_{3}$Au$_{3}$.

\begin{figure}[h!]
\centering
\includegraphics[width=\linewidth]{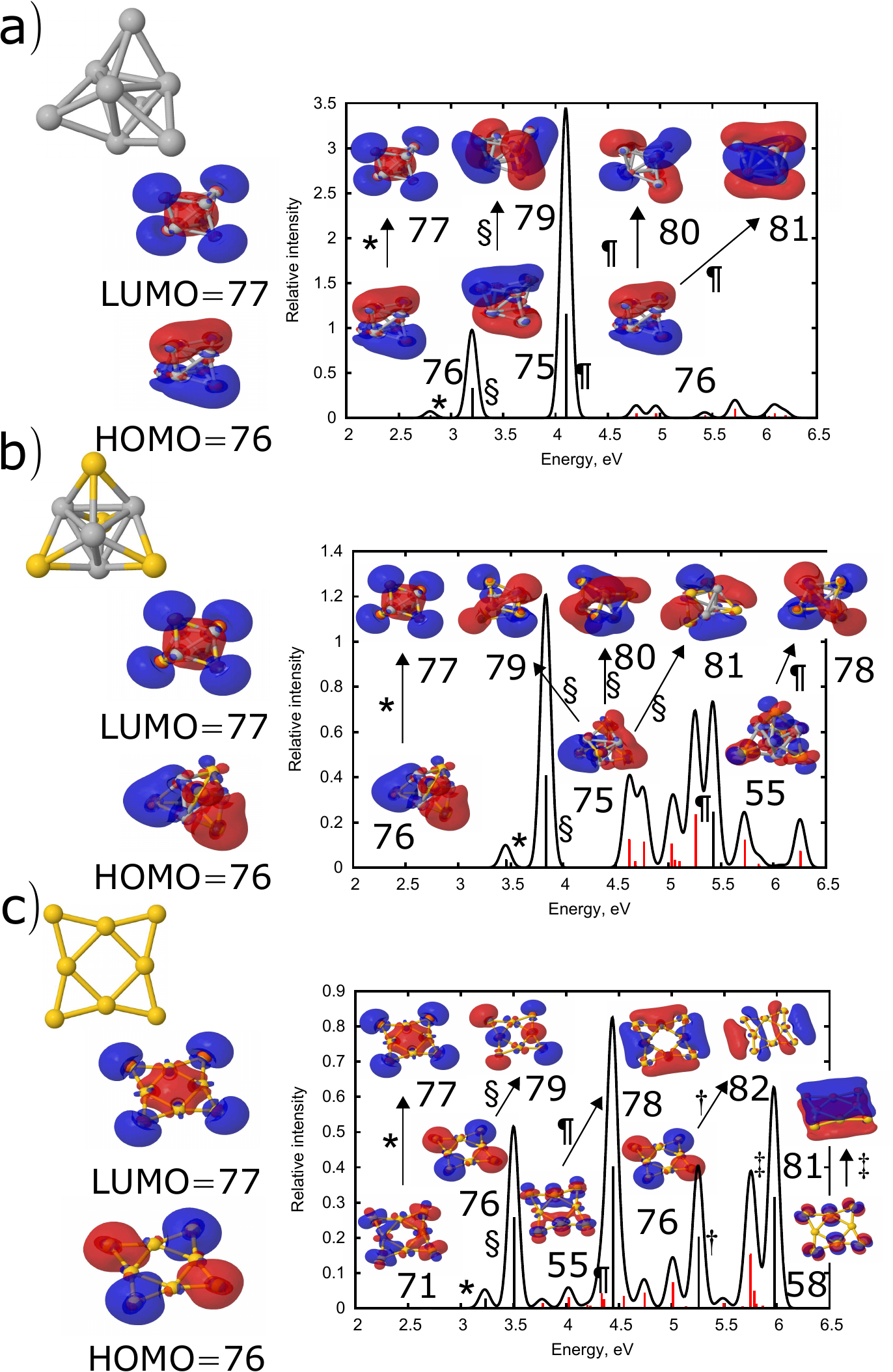}
\caption{Comparison of the excitation transitions for \mbox{(a) Ag$_{8}$}, (b) Ag$_{4}$Au$_{4}$, and (c) Au$_{8}$. See caption to Fig.~\ref{fig02} for details.}
\label{fig03}
\end{figure}

For larger mixed clusters the spectra soon become quite distinct and independent from that of the pure clusters, particularly when their structures are different from either of the pure clusters (i.e. Ag$_{5}$Au$_{5}$ and Ag$_{6}$Au$_{6}$, see Supplementary Information). Even for the 8-atom cluster, where Ag$_{4}$Au$_{4}$ and Ag$_{8}$ have the same geometries, there is little obvious similarity between the spectra (Fig.~\ref{fig03}). The low-energy peaks in the Ag$_{4}$Au$_{4}$ spectrum involve similar orbitals to those of Ag$_{8}$ spectrum, but in contrast to Ag$_{8}$, whose spectrum is virtually featureless above 4.2~eV, Ag$_{4}$Au$_{4}$ has a whole series of peaks in the high-energy region that involve transitions from low-lying orbitals (e.g. peak at 5.25 eV) in a manner more similar to the gold cluster.

Therefore, the analysis of the data above indicates that alloying results not just in a mixture of the features of pure clusters, but can also yield unique geometrical configurations and spectral properties. 

\section{Clusters stabilized by DNA bases}

\subsection{Role of base identity}

Several experimental studies have indicated that silver clusters can be stabilized by the interaction with cytosine-based oligomers,~\cite{doi:10.1021/ja031931o, doi:10.1021/jp0648487, doi:10.1021/jp804031v, C0CC05061J} or even be encircled by a DNA hairpin loop.~\cite{Oneill_Velazquez_2009, Oneill_Fygenson_2012, doi:10.1021/acs.analchem.5b03166, doi:10.1021/acs.jpcc.5b08834} For small gold clusters, on the other hand, preferential binding to purine bases has been predicted theoretically.~\cite{doi:10.1021/acs.jpcc.5b04948} In order to rationalize these observations and to provide further insights into possible mechanisms of binding, a reliable identification of the ground-state structures of the cluster-organic aggregates and a careful analysis of the chemical nature of the bonds formed would be particularly useful. Furthermore, a comparison of the complexes formed by a cluster with all possible nucleobases would also help to identify any special features of cytosine.

\begin{figure}[!ht]
\centering
\includegraphics[width=\linewidth]{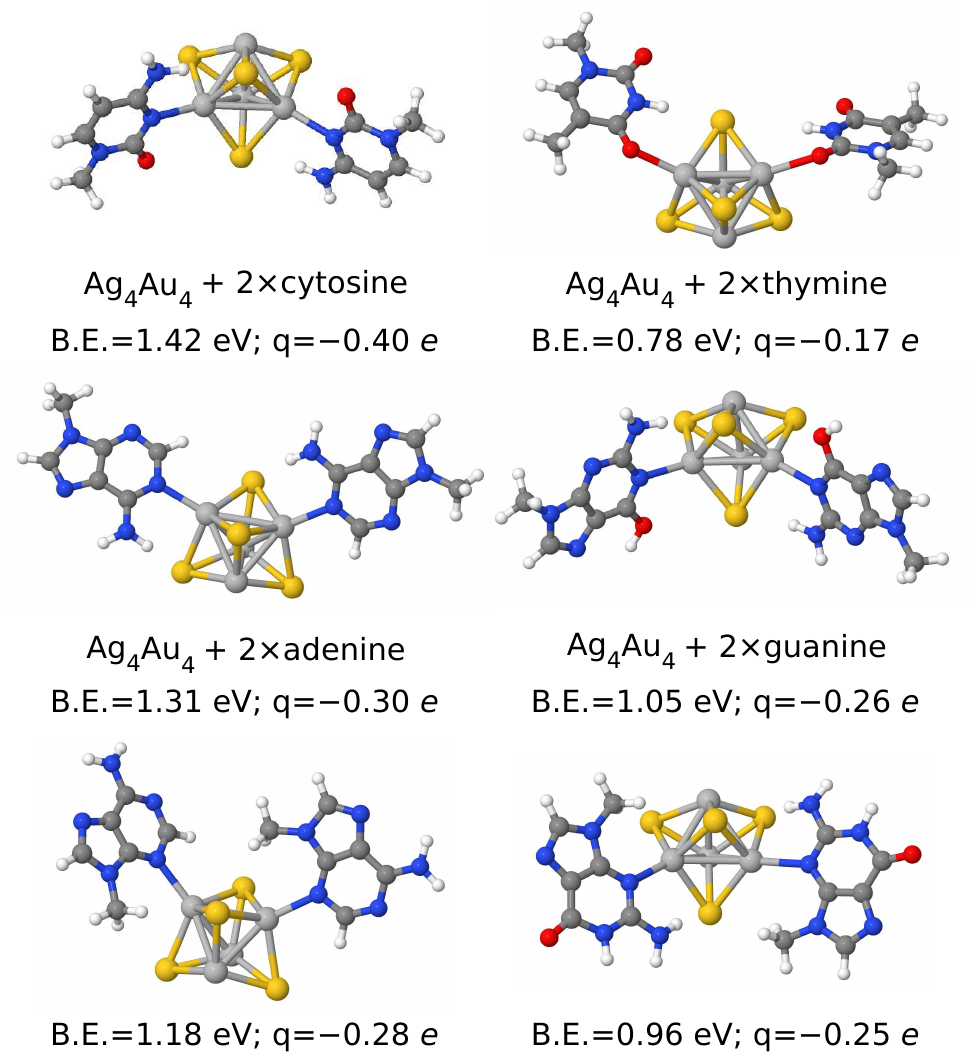}
\caption{Ground-state structures (as well as selected low-lying isomers of the adenine and guanine stabilized clusters) of the Ag$_{4}$Au$_{4}$ cluster with two identical bases. The numbers correspond to the binding energies and total Mulliken charges on the cluster.}
\label{fig04}
\end{figure}

We ran global geometry optimization on the intermediate size Ag$_{4}$Au$_{4}$ cluster with two identical bases, as a minimal example fragment of an embedded nanoalloy (Fig.~\ref{fig04}). Note that methylated bases are used in order to better mimic the configuration of the base in a real DNA fragment. Calculated binding energies indeed reveal that the Ag$_{4}$Au$_{4}$ cluster has the highest affinity towards two cytosines (1.42~eV), while two thymine bases exhibit much weaker cluster stabilization (0.78~eV). Our calculations indicate that pure silver clusters, pure gold clusters, and silver-gold nanoalloys prefer to bind to the unprotonated ring nitrogen atoms of the nucleobases. This can be explained by the basicity of such centers, which facilitates charge transfer to the attached cluster. The bond is formed due to the lone pair on the nitrogen atom interacting with the metal cluster. In the case of nanoalloys the bonds are formed with the less electronegative silver, not gold atoms, which is due to the partial positive charge accumulated on silver. For thymine, as no such nitrogen center is available, the bonding is instead mediated through a less basic lone pair on oxygen, and so is less energetically favourable. Therefore, cytosine, adenine, and guanine are potentially the most likely bases to stabilize clusters embedded into a DNA fragment. 

Two adenines yield an only slightly lower binding energy of 1.31~eV, compared to the 1.42~eV of the cytosine-based aggregate. Every adenine provides two unprotonated nitrogens on the 6-membered ring. Binding to the nitrogen between the CH and CNH$_{2}$ groups yields the binding energy of 1.31~eV, while binding to the other nitrogen atom results in a slightly lower value of 1.18~eV. Whether the second configuration would be sterically accessible in an actual DNA fragment where the CH$_{3}$ group is replaced by a ribose sugar connected to the DNA backbone is not clear. Steric inaccessibility is also the main reason we do not discuss binding to the nitrogen atoms of the 5-membered rings. Furthermore, although such configurations were identified in the global optimization, binding of the cluster to the 5-membered ring yields significantly lower binding energies of about $0.8$~eV.

Guanine has two nitrogens that are in the 6-membered ring: one protonated and one unprotonated. The unprotonated center yields only 0.96~eV binding energy towards Ag$_{4}$Au$_{4}$. In fact, it is more favourable (binding energy 1.05~eV) for guanine to bind in its enol tautomer where a hydrogen atom has been transferred from nitrogen to oxygen, despite the cost of this tautomerization (which is quite low at about $0.25 - 0.30$~eV~\cite{Satpati28102014}). It is noteworthy that the resulting unprotonated ring nitrogen has a local environment resembling that for cytosine. Although thymine can also form a similar tautomer with an unprotonated ring nitrogen atom, this isomer is much less stable,~\cite{thymine_tautomers_1, thymine_tautomers_2} and we were unable to identify any energetically favourable aggregates of that thymine form with Ag$_{4}$Au$_{4}$.

Thus, in the case of cytosine, adenine, and guanine we observe binding of the Ag$_{4}$Au$_{4}$ cluster to an unprotonated ring nitrogen, which is in line with previous studies,~\cite{Gwinn_TDDFT} with cytosine exhibiting stronger binding than both adenine and guanine. The relative binding energies reflect the relative basicity of the corresponding ring nitrogens,~\cite{DNA_bases_basicity} with higher basicity enabling greater charge transfer from the nucleobases to the metal cluster.










\subsection{Cytosine-stabilized clusters}\label{geometry_binding}

\begin{figure}[!ht]
\centering
\includegraphics[width=\linewidth]{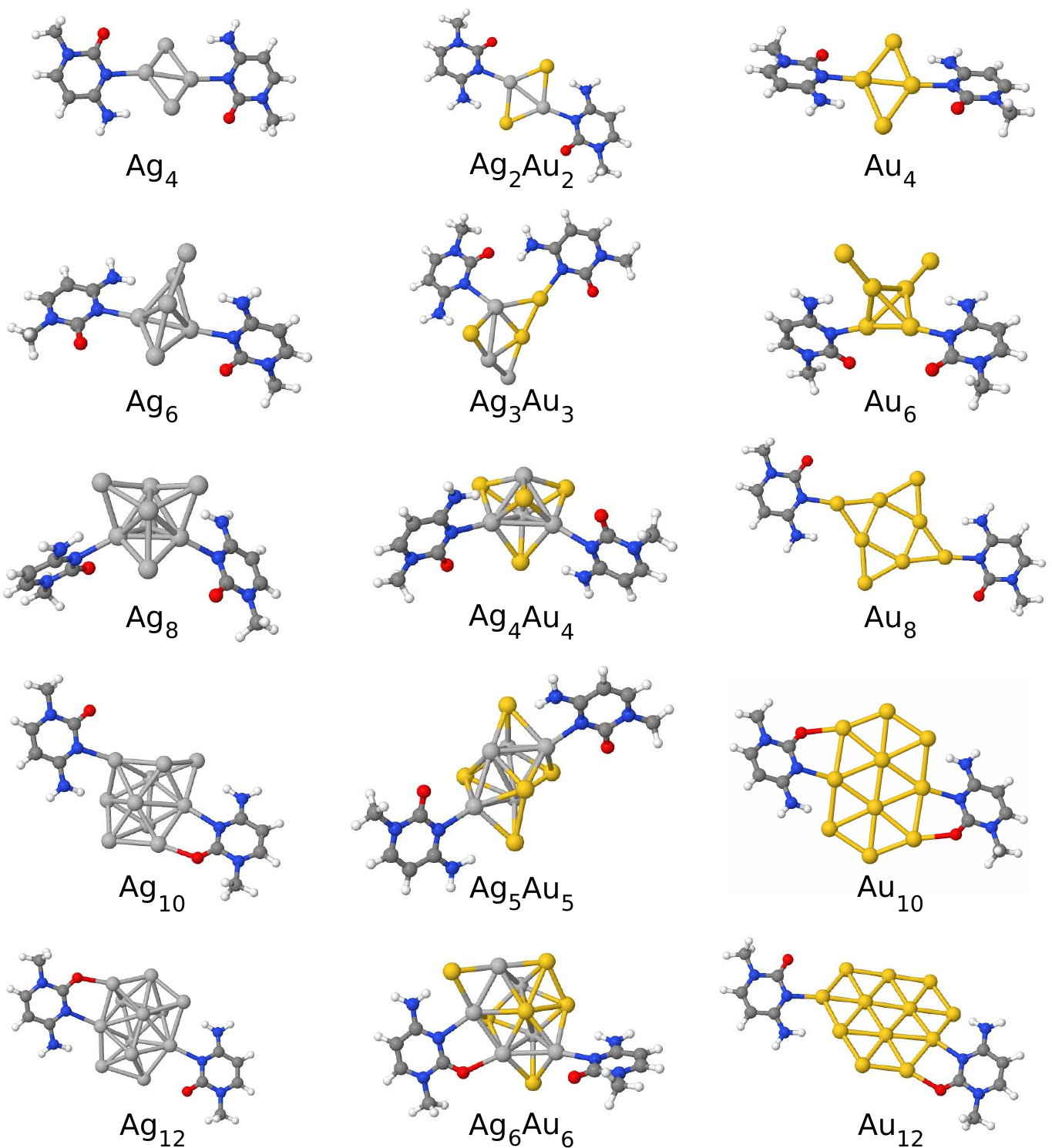}
\caption{Ground-state structures of pure and alloyed clusters with two cytosine bases.}
\label{fig05}
\end{figure}

Given the energetic preference of the Ag$_{4}$Au$_{4}$ cluster to bind to cytosine bases, we therefore performed global geometry optimization on all pure and alloyed clusters with two methylated cytosines (Fig.~\ref{fig05}). Ag$_{4}$, Ag$_{8}$, and Ag$_{12}$ retain their geometrical structures upon aggregation with cytosines, whereas Ag$_{6}$ no longer adopts a planar configuration and the original Ag$_{10}$ $C_{2}$ configuration gets distorted. Gold clusters are also able to mainly retain their structures with only Au$_{6}$ changing from planar into a more compact structure, while the other four clusters are still mostly unperturbed upon aggregation. Alloyed clusters exhibit more complex behaviour. While Ag$_{2}$Au$_{2}$ and Ag$_{4}$Au$_{4}$ retain their configuration, the other three clusters are perturbed. The most interesting example is Ag$_{3}$Au$_{3}$: the cluster is still nearly planar, but the chemical ordering of the atoms is changed: such a configuration is 0.08~eV more stable than the equivalent derived from the $D_{3h}$ isolated Ag$_{3}$Au$_{3}$ global minimum. It is also the only nanoalloy where gold atoms directly interact with the bases: the rest of the clusters only bind to the cytosines via silver.

\begin{table*}[htp!]
\centering
\begin{tabular}{m{0.55cm} c c | m{1.10cm} c c | m{0.55cm} c c}

\hline
 Ag$_{n}$     & BE/eV & q/$e$ &  Ag$_{n}$Au$_{n}$ & BE/eV &  q/$e$ &  Au$_{n}$ & BE/eV & q/$e$ \\
\hline 

Ag$_{4}$  & 1.79 & $-$0.32 & Ag$_{2}$Au$_{2}$  & 1.92 & $-$0.34  & Au$_{4}$  & 2.46 & $-$0.39  \\

Ag$_{6}$  & 1.13 & $-$0.36 & Ag$_{3}$Au$_{3}$  & 1.23 & $-$0.36  & Au$_{6}$  & 1.59 & $-$0.48  \\

Ag$_{8}$  & 1.06 & $-$0.47 & Ag$_{4}$Au$_{4}$  & 1.42 & $-$0.40  & Au$_{8}$  & 2.26 & $-$0.43  \\

Ag$_{10}$ & 1.32 & $-$0.44 & Ag$_{5}$Au$_{5}$ & 1.66  & $-$0.52  & Au$_{10}$ & 1.75 & $-$0.71  \\

Ag$_{12}$ & 1.34 & $-$0.51 & Ag$_{6}$Au$_{6}$ & 1.42  & $-$0.52  & Au$_{12}$ & 1.89 & $-$0.62  \\

\hline


\end{tabular}
\caption{Binding energies for pure and alloyed clusters with two cytosine bases, and total Mulliken charge on the cluster, illustrating charge transfer towards the cluster from the base.}
\label{table_2}
\end{table*}

The binding energies and the total Mulliken charges on the clusters are presented in Table~\ref{table_2}. In all cases we again find that the clusters become partially negatively charged due to charge transfer from the cytosines. We also find that for each cluster type the largest binding energy belongs to the smallest 4-atom cluster. The smallest clusters are least stable individually, but most stabilized by association with the nucleobases. Gold clusters, being more electronegative, exhibit larger binding energies, which can also be seen in the ability of gold atoms to form additional Au--O bonds and \mbox{N--H$\cdots$Au} hydrogen bonds, along with the more conventional for nucleobases Au--N bonds.~\cite{doi:10.1021/jp054708h, C3CS60251F} This is also reflected in the geometrical configurations of larger gold clusters: in the case of Au$_{10}$ and Au$_{12}$, both N and O can interact with the cluster, which causes the cytosines to be oriented coplanar with the cluster. 

Gold clusters also have the largest amount of charge transfer from the nucleobase to the cluster, which is an important aspect of the cytosine binding. One can see a general trend of the amount of charge transferred towards cluster increasing from silver clusters through nanoalloys to gold clusters, which stems from the greater electronegativity of the gold atoms. Reflecting this, the low-lying unoccupied orbitals of the gold clusters are lower in energy than those of the silver clusters, making it more favourable for them to receive additional negative charge. In the case of alloy clusters, cytosine tends to bind to the surface silver atoms, which are the atoms with the partial positive charge. Interestingly, the silver atoms directly connected to the nitrogen atoms of cytosine increase their positive partial charge in the case of nanoalloys, while their more electronegative neighbouring gold atoms host the excess negative charge. For example, in the case of Ag$_{4}$Au$_{4}$, the two silver atoms connected to nitrogen have a $+0.52$\,$e$ charge (compared to $+0.43$\,$e$ in the bare cluster), while the neighbouring gold atoms have $-0.59$\,$e$ each (compared to $-0.43$\,$e$ in the bare cluster). Bimetallic clusters generally have an intermediate binding energy and charge transfer values between the corresponding pure silver and gold clusters. This indicates that alloying can be used to tune the binding strength between the cluster and the DNA fragment.

\begin{figure}[!ht]
\centering
\includegraphics[width=\linewidth]{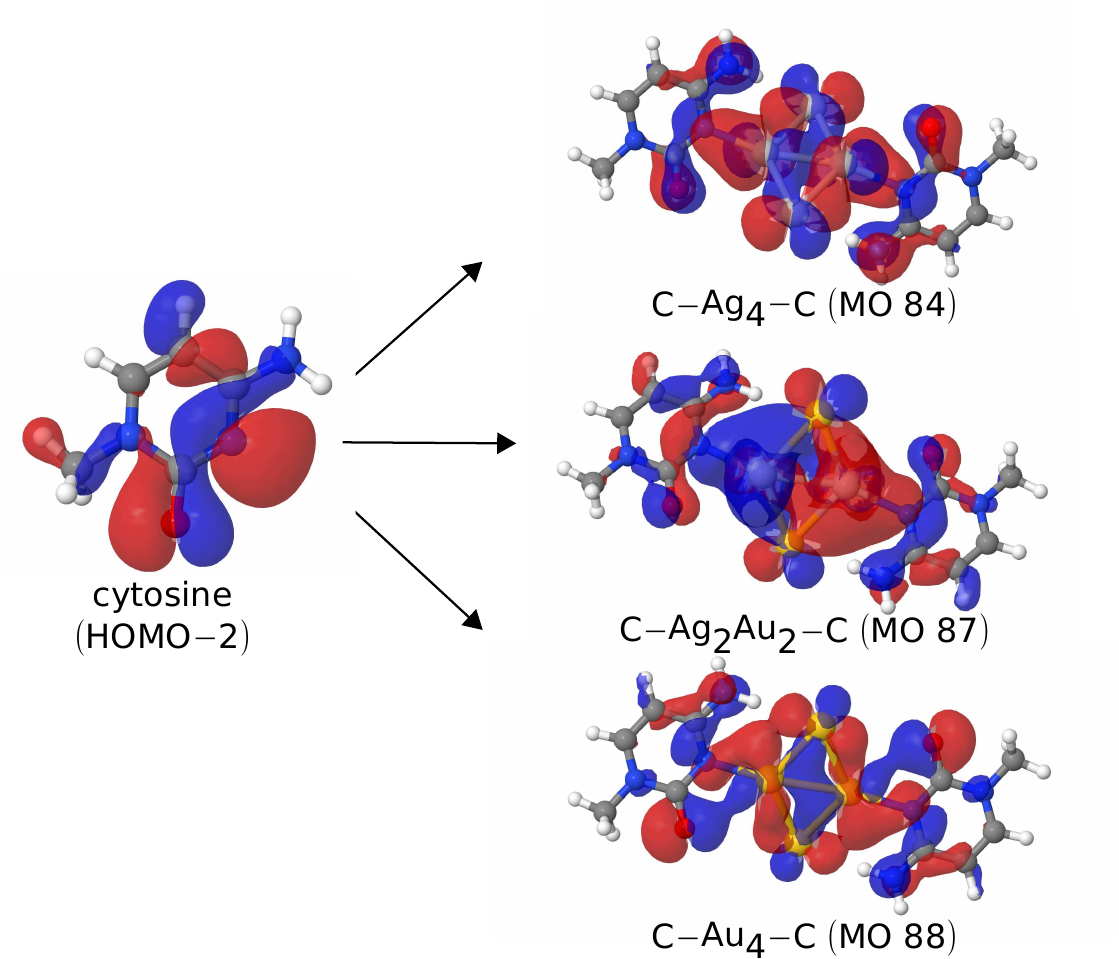}
\caption{The binding of cytosine to Ag$_{4}$, Ag$_{2}$Au$_{2}$, and Au$_{4}$. The lone pair of the unprotonated ring N atom of cytosine (populating the HOMO$-2$ orbital) facilitates the binding with the metal clusters to form low-lying bonding orbitals (HOMO=104).}
\label{fig06}
\end{figure}

Fig.~\ref{fig06} illustrates the binding of two methylated cytosines to the smallest four-atom clusters. The lone pair on the ring nitrogen atom of cytosine is responsible for the formation of some of the lower-lying bonding orbitals in the C--cluster--C aggregates. In the case of the bimetallic Ag$_{2}$Au$_{2}$ cluster, the lone pair from the unprotonated ring nitrogen atom of the cytosine clearly hybridizes with the cluster's LUMO. In the case of Au$_{4}$, also a lone pair on the oxygen is able to participate in bonding interactions with the nearest gold atom. This is in line with the observed higher binding energy for gold clusters. The bonding to the pure silver Ag$_{4}$ cluster is again facilitated exclusively via a lone pair of nitrogen, although the lower binding energy indicates a weaker interaction.

It is also noteworthy that visual analysis of the occupied orbitals (e.g. as seen in Figs.~\ref{fig07} and~\ref{fig08}) reveals higher level of hybridization for the clusters with higher binding energies towards cytosines, most notably for pure gold clusters. Silver clusters exhibit much lower hybridization with the organic fragments; however, in the nanoalloys, silver atoms are usually the preferred contact to the base due to the partial positive charge, tunneling the excess negative charge towards gold atoms. In such a manner, both strong binding and conservation of the main geometrical and optical properties of a cluster can be achieved, which illustrates the complementary properties of the silver and gold atoms in a nanoalloy.

\subsection{Absorption spectra}

\begin{figure}[t]
\centering
\includegraphics[width=\linewidth]{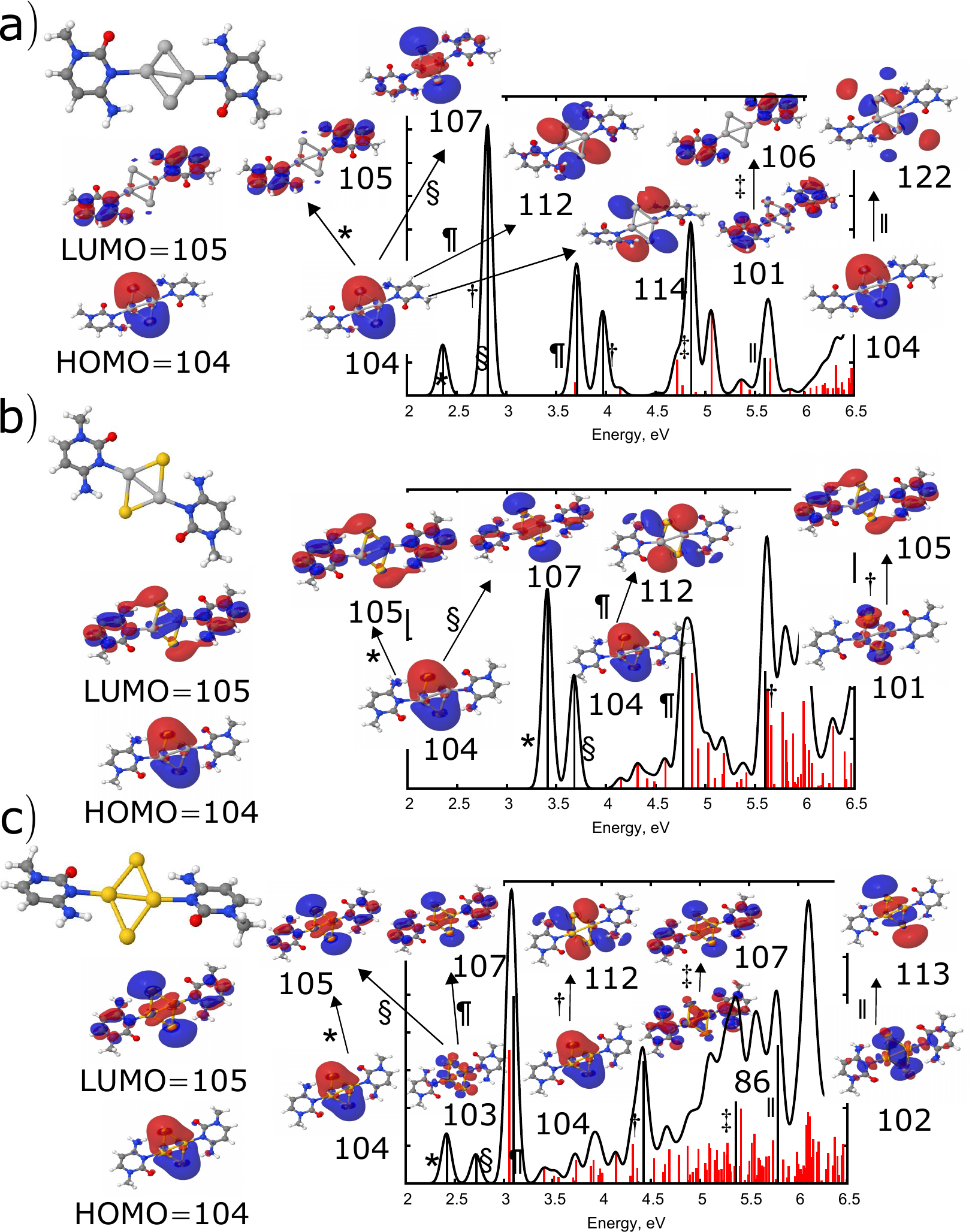}
\caption{Comparison of the excitation transitions for (a) Ag$_{4}$, (b) Ag$_{2}$Au$_{2}$, and (c) Au$_{4}$ clusters associated with two cytosine bases. See caption to Fig.~\ref{fig02} for details.}
\label{fig07}
\end{figure}

As with their geometries, the absorption spectra of Ag$_{4}$, Au$_{4}$, and Ag$_{2}$Au$_{2}$ retain some of their main features on association. The total number of peaks increases due to the involvement of the electrons on the cytosines. As well as the transitions that predominantly involve the electrons on the metal cluster and are present in the spectra of the isolated clusters, there are additional bands involving electron transfer between the metal cluster and the bases, and  transitions that mainly involve the electrons on the bases leading to more complex spectra than for the bare clusters. For all three complexes, the low energy transitions are dominated by transitions from the HOMO, which has the same character in all three systems and is the same as for the isolated clusters. The LUMOs of all three clusters, on the other hand, show substantial hybridization between the metal cluster and the cytosine fragments (Fig.~\ref{fig07}). 

In the case of Ag$_{4}$, this leads to a new low-energy band at 2.4~eV, corresponding to the transition from the cluster-centered HOMO to the cytosine-centered LUMO. Peaks at 2.8~eV, 3.7~eV, and 4.0~eV are due to the metal cluster and therefore conserved, although slightly shifted to the lower energies. In the case of Ag$_{2}$Au$_{2}$, the first two bands at 3.45~eV and 3.6~eV involve transitions from the cluster-centered HOMO to the hybrid LUMO and LUMO$+2$, respectively. Although in the bare Ag$_{2}$Au$_{2}$ cluster there is only one prominent peak in this energy region corresponding to HOMO$\rightarrow$LUMO$+1$, its position (3.5 eV) and the character of the acceptor orbital is very similar to the cluster stabilized with two cytosines, where, due to mixing, there are two nearby orbitals with very similar character for the metallic part. In the higher-energy regions one can see some transitions analogous to the bare clusters, but the lines become more numerous making the spectrum more complex. Similarly to Ag$_{2}$Au$_{2}$, in the case of cytosine-stabilized Au$_{4}$ the first HOMO$\rightarrow$LUMO band at 2.4~eV involves the same metal cluster orbitals as the 3.1~eV HOMO$\rightarrow$LUMO$+1$ peak of the bare Au$_{4}$ cluster. Likewise, the 4.45~eV band of the cluster/cytosine aggregate resembles the peak at 4.7~eV of the bare cluster. The higher-energy region, however, reflects the gold cluster's propensity to stronger hybridization with the cytosine molecules, with many excitation transitions that bear little similarity with the bare Au$_{4}$ cluster. 

\begin{figure}[h]
\centering
\includegraphics[width=\linewidth]{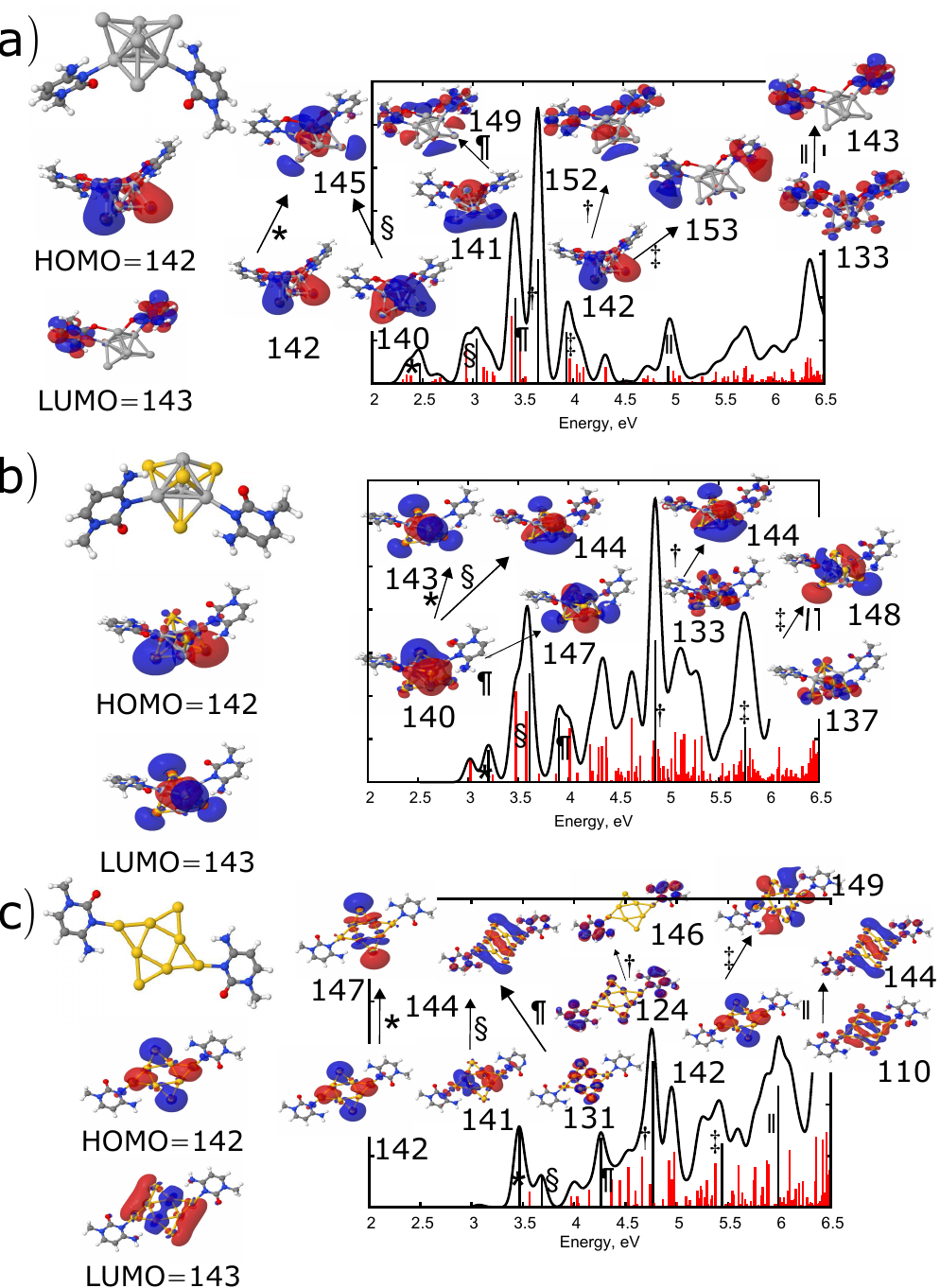}
\caption{Comparison of the excitation transitions for (a) Ag$_{8}$, (b) Ag$_{4}$Au$_{4}$, and (c) Au$_{8}$ clusters associated with two cytosine bases. See caption to Fig.~\ref{fig02} for details.}
\label{fig08}
\end{figure}

Stronger geometrical distortions yield larger differences between the absorption spectra of the individual clusters and the clusters stabilized with cytosines (Ag$_{3}$Au$_{3}$), as does increasing system size (Ag$_{5}$Au$_{5}$ and Ag$_{6}$Au$_{6}$, see Supplementary Information). The absorption spectra become more complex, and the number and intensity of additional high energy transitions rise. The level of hybridization of the orbitals involved in the transitions is also noticeably higher for larger clusters. This trend of increasing complexity in the high-energy region is also evident for the Ag$_{8}$, Ag$_{4}$Au$_{4}$, and Au$_{8}$ (Fig.~\ref{fig08}). However, in the case of Ag$_{4}$Au$_{4}$ one can still identify many of the original excitation transitions, with the orbitals involved mostly formed by the $d$-electrons of the cluster constituent metals. For instance, the HOMO and LUMO of the bare cluster retain their character upon aggregation with cytosines. In this respect, Ag$_{4}$Au$_{4}$ still shows a certain robustness of its properties even upon aggregation with two cytosine molecules.

\section{Towards hairpin embedding}

\begin{figure}[h!]
\centering
\includegraphics[width=0.6\linewidth]{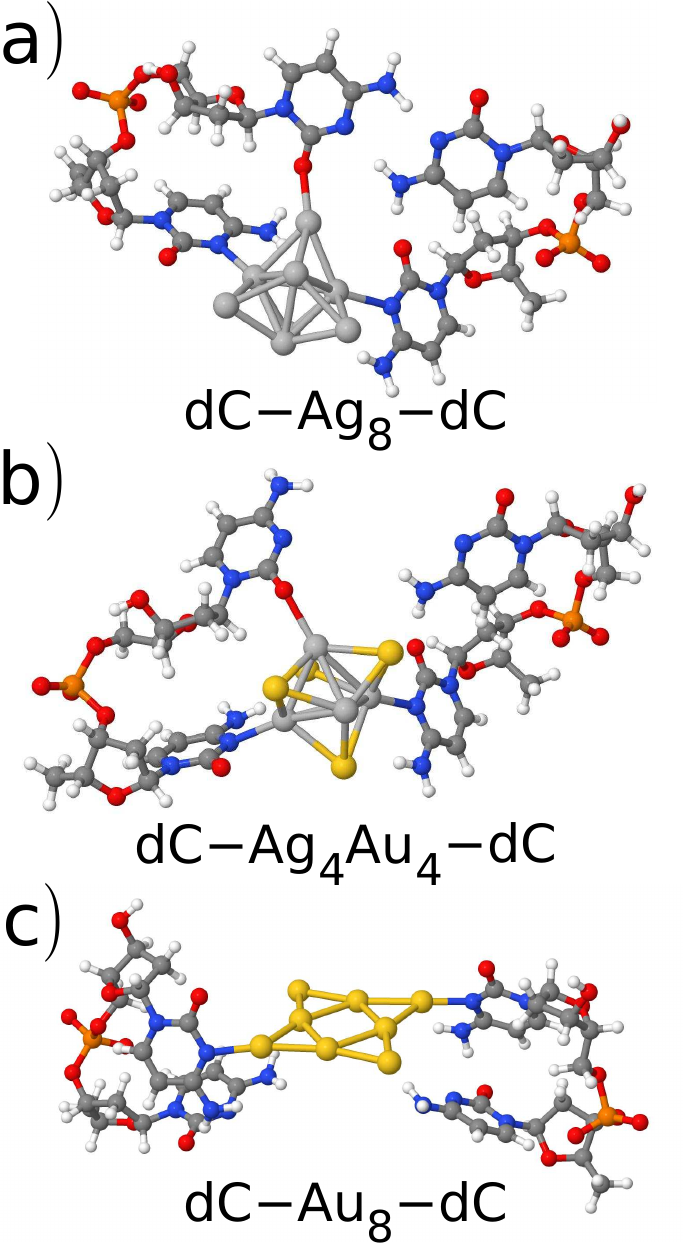}
\caption{(a) Ag$_{8}$, (b) Ag$_{4}$Au$_{4}$, and (c) Au$_{8}$ stabilized with two cytosine dinucleotides (dC). Each structure corresponds to the lowest configuration found in a series of local optimizations.}
\label{fig09}
\end{figure}

Although two cytosine bases can affect the geometrical structure and absorption spectra of small clusters, many of the main geometrical and spectral features persist in most of the Ag--Au nanoalloys, allowing the construction of biological fragments with optical properties similar to those of individual bimetallic clusters. Is it thus possible to embed such clusters into a larger DNA-based structure, closer in size to, say, the hairpin suggested in Ref.~\citenum{Oneill_Fygenson_2012}? As the next step towards hairpin embedding, we have chosen two cytosine dinucleotides as a model fragment of the hairpin sufficient to encapsulate an eight-atom cluster. Ag$_{8}$, Ag$_{4}$Au$_{4}$, and Au$_{8}$ were chosen on the basis of their stability and clear spectral features. The optimized aggregates, identified as the lowest-energy structures found by the local optimization of ten different initial configurations, are presented in Fig.~\ref{fig09}. Although the identified structures are unlikely to be the true global minima, they are likely to be low-lying isomers that are representative of the process of aggregation of the pre-formed clusters with DNA fragments. As can be seen in Fig.~\ref{fig09}, both tetrahedral (Ag$_{8}$, Ag$_{4}$Au$_{4}$) and planar (Au$_{8}$) clusters are able to retain their original configurations. In all cases, enlarged DNA fragments allow efficient association of the metal cluster with the cytosine dinucleotides by providing additional interactions compared to the cluster stabilized with two individual cytosine bases (Fig.~\ref{fig05}). This is also reflected in higher binding energies of 1.98~eV, 1.67~eV, and 3.01~eV for Ag$_{8}$, Ag$_{4}$Au$_{4}$, and Au$_{8}$, respectively, compared to 1.06~eV, 1.42~eV, and 2.26~eV for smaller complexes. The charge transfer towards the cluster is also intensified in all cases, with the excess negative charge on the cluster increased to $-0.62$\,$e$, $-0.68$\,$e$ and $-0.70$\,$e$ for the dinucleotide-stabilized Ag$_{8}$, Ag$_{4}$Au$_{4}$, and Au$_{8}$, respectively. 

Comparison of the optical absorption spectra of Ag$_{8}$, Au$_{8}$, and Ag$_{4}$Au$_{4}$ clusters aggregated with two cytosines and with two cytosine dinucleotides reveals that the main spectral features are still due to the orbitals centered on the metal clusters, although more complex ligands lead to an increase in the number of transitions in the higher energy region of the spectra (see Supplementary Information). For instance, in the case of the Ag$_{4}$Au$_{4}$ cluster all the orbitals involved in the transitions below 5.5~eV are localized on the metal atoms, while only some of the higher energy excitations are due to the cytosine dinucleotides.

\begin{figure}[h!]
\centering
\includegraphics[width=0.7\linewidth]{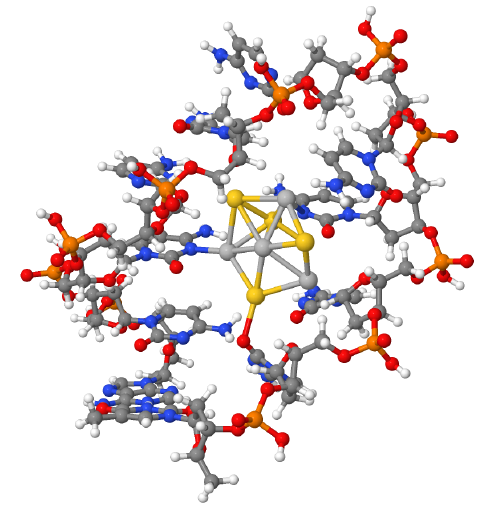}
\caption{Locally optimized Ag$_{4}$Au$_{4}$ embedded into a DNA hairpin with 9 cytosines in the loop.}
\label{fig10}
\end{figure}

Thus compact bimetallic clusters are able to retain their main properties upon aggregation with larger DNA fragments with alloying allowing tuning of the binding energies, the charge transfer, and the optical properties. Can such clusters be embedded into a full-size hairpin and still retain their properties? To simulate such embedding, we insert a tetrahedral global minimum configuration of the Ag$_{4}$Au$_{4}$ bimetallic cluster into a pre-optimized cytosine hairpin, consisting of 9 cytosine bases in the loop and one adenine-thymine base pair in the stem, by placing the cluster between the inward-facing bases, as proposed in Ref.~\citenum{Oneill_Fygenson_2012}, and relax the obtained initial structure locally. While such an approach will obviously not lead to a globally optimal configuration, it can still facilitate a proof-of-principle check on whether a compact cluster structure is able to retain its stability upon the insertion into a hairpin, and whether such aggregation is energetically favourable. Fig.~\ref{fig10} depicts a locally optimized aggregate of a cluster embedded into a cytosine-based  hairpin. This example proves that, although quite distorted, the cluster is able to retain a compact geometry and overall stability. The binding energy of 3.23~eV (with respect to the individual ground states of the Ag$_{4}$Au$_{4}$ cluster and the hairpin fragment) is larger than those of the other considered clusters stabilized by smaller organic fragments, including pure gold clusters. Of course, it should be kept in mind that this is a locally optimized example structure, and therefore other structures with potentially stronger binding would be expected.

In order to check whether non-compact cluster geometries could be stabilized by a hairpin, we also embedded an energetically higher-lying elongated Ag$_{4}$Au$_{4}$ isomer, identified during the global geometry sampling. During the local optimization of this rod-shaped Ag$_{4}$Au$_{4}$ cluster embedded into the hairpin, the cluster underwent significant distortion, breaking up into two fragments (see Supplementary Information), and overall the aggregate turned out to be energetically unfavourable. An elongated pure silver cluster is even less likely to lead to a thermodynamically stable configuration. In the case of Ag$_{8}$, the cluster loses two atoms, which are then used to mediate a direct contact between the bases, thus ``stitching'' the hairpin structure (see Supplementary Information). While the formation of such single-atom bridges at the centre of a DNA duplex has been suggested for silver ions,~\cite{Schutz_Gwinn_2013, Copp_Gwinn_2014, Swasey_Gwinn_2014, Fortino_Russo_2015} such a configuration is energetically unfavourable for the neutral cluster embedded in the hairpin. Inserting Ag$_{4}$Au$_{4}$ into a hairpin, on the other hand, allows retaining the compact structure of the cluster and is energetically favourable, suggesting the potential experimental feasibility of constructing such hybrid cluster/DNA compounds.

\section{Conclusions}

In summary, we have systematically studied the geometric structures and optical properties of the Ag--Au nanoalloys, both as individual clusters and stabilized with DNA fragments. We have shown that most of the small nanoalloys retain the geometries of either of their ``parent'' clusters. An important effect of alloying, however, is a partial charge transfer within the cluster from silver to more electronegative gold atoms. For the smaller nanoalloy clusters the optical absorption spectra appear to be mainly composed of the transitions analogous to those of the ``parent'' clusters, with the excitations in the lower energy region corresponding to silver, and larger contribution of gold in the higher energy region. For the larger clusters, however, new transitions are observed due to changes in cluster geometries, molecular orbital types, and relative orbital energies, with alloying thus directly influencing the optical properties.

Bimetallic clusters form the most stable aggregates with cytosine bases, which can be explained by the basicity of the unprotonated ring nitrogen atoms that act as the preferred binding sites. The cytosine-metal bond is formed due to the lone pair on the nitrogen atom interacting with the metal cluster. Perhaps somewhat counterintuitively, given that cytosine binds most strongly to pure gold clusters, the bonds between the cytosine and the nanoalloy are preferentially formed with the less electronegative silver, not gold atoms; this is due to the partial positive charge accumulated on silver. Moreover, the silver atoms directly connected to the nitrogen atoms of cytosine increase their positive partial charge in the case of nanoalloys, while their more electronegative neighbouring gold atoms host the excess negative charge. It is indicative of the complementary properties of the silver and gold atoms, ensuring effective binding towards DNA fragments while preserving integrity of the cluster properties. 

Many of the spectral features are conserved upon aggregation with two cytosine bases. The overall complexity of the absorption spectra, however, increases due to the newly emerged transitions involving orbitals with major contributions from the organic fragments. This already renders alloying a suitable tool for adjusting the level of interaction between the metal atoms and the organic fragments, as well as the nature of the orbitals taking part in excitation transitions, thus allowing direct tuning of the optical properties.  

Finally, the optimized structure of the Ag$_{4}$Au$_{4}$ cluster embedded into a cytosine-based 9-nucleotide hairpin loop indicates that such clusters can retain overall stability and compact structures upon aggregation with large organic fragments. Such stability suggests the potential experimental feasibility of assembling hybrid cluster-based optical materials relevant for biological and medical applications.

\section*{Computational details}

All local geometry optimizations of the discussed structures, and subsequent electronic structure analysis were carried out with the plane-wave density functional theory (DFT) package \texttt{CASTEP}.\cite{CASTEP_reference} Electronic exchange and correlation was treated within the generalized-gradient approximation functional due to Perdew, Burke and Ernzerhof (PBE).\cite{PBE_reference} The core electrons were described using ultrasoft pseudopotentials, whereas the valence electrons were treated with a plane-wave basis set with a cut-off energy of 400\;eV. Local structure optimization is done using the Broyden-Fletcher-Goldfarb-Shanno method,\cite{Nocedal_Wrigth_Num_Opt} relaxing all force components to smaller than 0.01~eV/{\AA}.

To obtain the ground-state structures for the smaller systems considered (individual clusters, and clusters aggregated with two bases), we relied on basin-hopping (BH) based global geometry optimization,\cite{Doye_1997, Wales_2000} using the DFT total energies and atomic forces calculated by \texttt{CASTEP}\cite{CASTEP_reference} as implemented in the \texttt{Atomic Simulation Environment (ASE)} suite.\cite{ASE} As global optimization of the larger structures (complexes with cytosine dinucleotides and a hairpin loop) is at the edge of the current computational capabilities, we relied on local optimization of several chemically-sensible initial configurations. While the structures presented here are thus unlikely to be true global minima, they represent typical, if not necessarily optimal, geometries for these compounds.

All geometries reported here correspond to the structures in vacuum. Considering the typical preparation methods of the cluster-DNA aggregates,~\cite{Buceta_2015, doi:10.1021/acs.analchem.5b01265} one would want to extend these results to clusters functioning in solution. However, we do not expect the general trends outlined here to change, as neither the nature of bonding nor the optical excitation transitions should be dramatically influenced by the solvent.~\cite{doi:10.1021/jp9051853, C4CP06103A}

To simulate the optical absorption spectra the first 400 excitation transitions were calculated with the \texttt{Gaussian~09} package~\cite{g09} using the long range corrected cam-b3lyp functional~\cite{cam-b3lyp} with lanl2dz basis set~\cite{lanl2dz} within the time-dependent density functional theory (TD-DFT)~\cite{TDDFT} approach. This approach has been validated by a good comparison of absorption spectra of pure clusters to experiment (see Supplementary Information, section SI).

Tkatchenko-Scheffler dispersion correction~\cite{TS_vdw} has been used to test the influence of the van der Waals energies arising from the attraction between induced dipoles formed due to charge fluctuations in the interacting species. We found that he dispersion correction shifts all energies more or less systematically to lower values by 0.1--0.3\,eV (see Supplementary Information). Subsequently, the qualitative picture presented in the manuscript does not change.

\section*{Acknowledgements}

Funding from the Engineering and Physical Sciences Research Council (project EP/J011185/1 ``TOUCAN: Towards an Understanding of Catalysis on Nanoalloys'') is gratefully acknowledged. Calculations were carried out using the ARCHER UK National Supercomputing Service (\url{www.archer.ac.uk}) and local compute facilities of the University of Oxford Advanced Research Computing (\url{http://dx.doi.org/10.5281/zenodo.22558}).

\section*{Supplementary Information}

See Supplementary Information for optical absorption spectra of individual and cytosine-stabilized pristine (Ag$_{4}$, Ag$_{6}$, Ag$_{8}$, Ag$_{10}$, Ag$_{12}$, Au$_{4}$, Au$_{6}$, Au$_{8}$, Au$_{10}$, Au$_{12}$) and alloyed (Ag$_{2}$Au$_{2}$, Ag$_{3}$Au$_{3}$, Ag$_{4}$Au$_{4}$, Ag$_{5}$Au$_{5}$, Ag$_{6}$Au$_{6}$) clusters, as well as for the relative stabilities of cluster-DNA aggregates. 

\bibliography{references}

\clearpage

\renewcommand{\thefigure}{Graphics}
\renewcommand{\figurename}{Table of Contents}
\begin{figure*}[h!]
\centering
\includegraphics[width=0.67\textwidth]{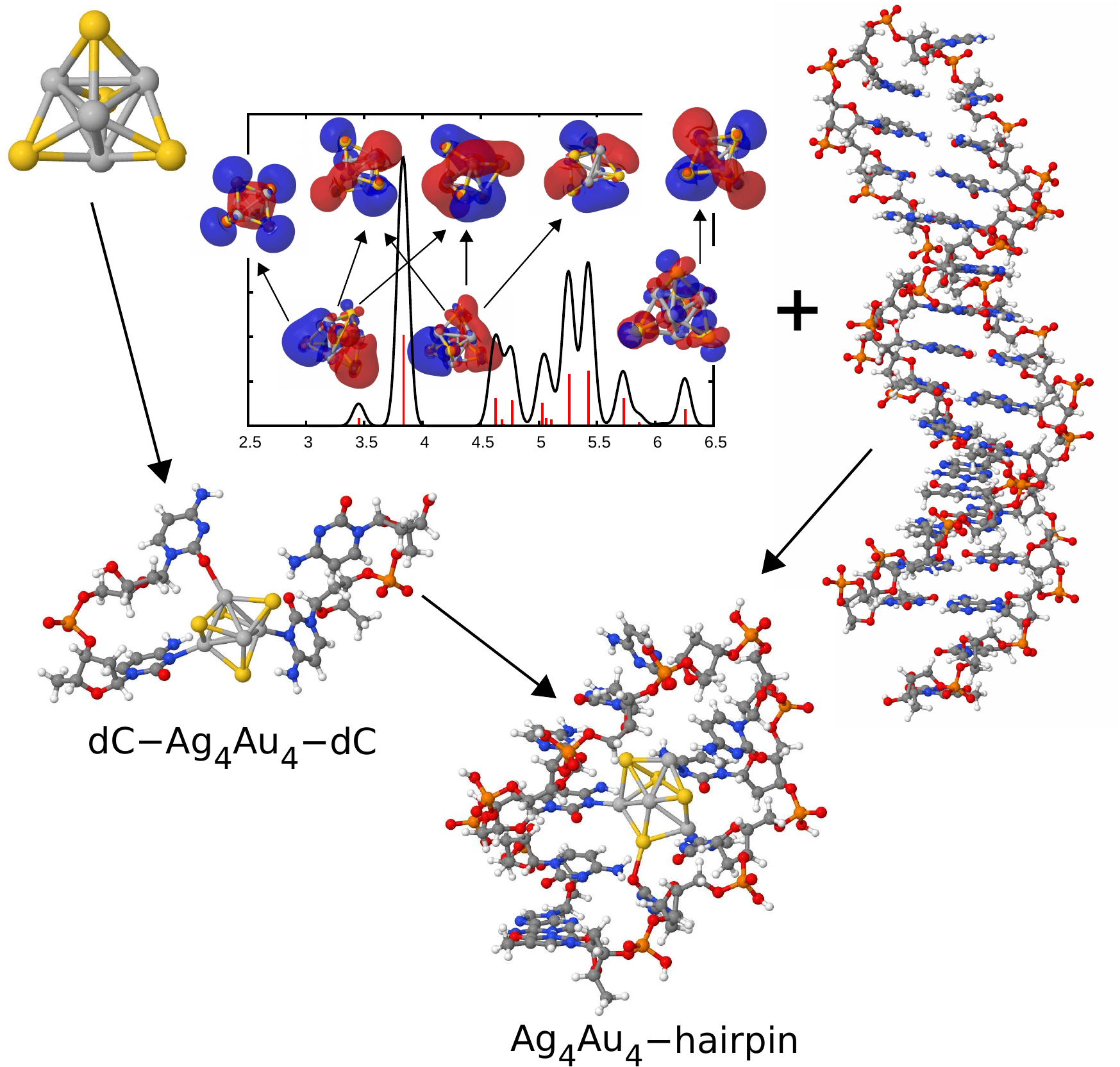}
\caption{}
\label{TOC}
\end{figure*}

\end{document}